\documentclass[useAMS,usenatbib]{mn2e}

\usepackage{graphicx}
\usepackage{longtable}

\title[Apsidal motion in HD~165052]{Apsidal motion in massive close binary systems.\\
I. HD~165052 an extreme case? 
\thanks{This work is based on observations made with three facilities: 
the J. Sahade telescope at Complejo Astron\'omico El Leoncito (CASLEO),
the du Pont telescope at Las Campanas Observatory (LCO), and
the 2.2-m telescope at La Silla Observatory (ESO) under programmes ID 083.D-0589 and 087.D-0946
}
}
\author[G. Ferrero, R. Gamen, O. G. Benvenuto and E. Fern\'andez-Laj\'us]{G. Ferrero$^{1,2}$\thanks{E-mail:
gferrero@fcaglp.unlp.edu.ar}\thanks{Visiting Astronomer, CASLEO (operated under
agreement between the Consejo Nacional de Investigaciones Cient\'{\i}ficas y
T\'ecnicas de la Rep\'ublica Argentina and the National Universities of La
Plata, C\'ordoba and San Juan).},
R. Gamen$^{1,2}$\thanks{Visiting Astronomer, LCO.}, 
O. Benvenuto$^{1,2}$\thanks{Member of the Carrera del Investigador Cient\'{\i}fico, Comisi\'on de Investigaciones Cient\'{\i}ficas de la Provincia
de Buenos Aires, La Plata, Argentina.} and E. Fern\'andez-Laj\'us$^{1,2}$\\
$^{1}$Facultad de Ciencias Astron\'omicas y Geof\'{\i}sicas. Universidad Nacional de La Plata,
Paseo del Bosque S/N, (1900) La Plata, Argentina\\
$^{2}$Instituto de Astrof\'{\i}sica de La Plata, CCT La Plata-CONICET ,UNLP, Argentina}

\begin{document}

\date{Accepted 2013 May 7.  Received 2013 May 4; in original form 2013 March 11}

\pagerange{\pageref{firstpage}--\pageref{lastpage}} \pubyear{201x}

\maketitle

\label{firstpage}

\begin{abstract}
We present a new set of radial-velocity measurements of the spectroscopic binary HD~165052
obtained by disentangling of high-resolution optical spectra. The longitude of the periastron 
($\varpi=60 \pm 2$ degrees) shows a variation with respect to 
previous studies. We have determined the apsidal motion
rate of the system $\dot{\varpi}=12.1 \pm 0.3 \degr$yr$^{-1}$, which was used to calculate
the absolute masses of the binary components: $M_1=~22.5 \pm 1.0~M_{\sun}$ and 
$M_2=20.5 \pm 0.9~M_{\sun}$. 
Analysing the separated spectra we have re-classified the
components as O7V\textit{z} and O7.5V\textit{z} stars.
\end{abstract}


\section{Introduction}
\label{intro}

O-type stars are objects that deserve great astrophysical interest.
Usually, they are located in star forming regions where, due to their high-velocity winds and intense UV radiation, they determine the dynamical and ionization state of the neighboring interstellar medium.
The turbulence they generate in the interstellar medium drives galactic dynamos. 
Furthermore, their high luminosities dominate the radiation of their birth regions, 
open clusters, spiral arms and even entire galaxies.
Besides, by mean of their explosive end as supernovae, 
they play a key role in the chemical evolution of the interstellar medium
and therefore in the metal enrichment of the successive generations of stars. 
However, several features of O-type stars
are still poorly known. Specially their masses represent a fundamental astrophysical parameter, which  
is still highly uncertain for various spectral sub-types. For this reason, to establish constraints 
to the masses of O-type stars is a very important task.  
The most direct and reliable method to determine stellar masses is the analysis of the Keplerian motion in
detached double-lined eclipsing binary systems.
The photometric and spectroscopic observation of such systems, and the subsequent analysis of the
variations in their brightness and radial velocities (RV), allows to calculate the parameters of
their orbital motion and the minimum masses of their components.
Furthermore, the observation of eclipses indicates that the orbital plane
lies close to the line of sight (i. e. its inclination $i$ is close to 90\degr). 
Then, using adequated models of stellar structure, it is possible to
derive absolute stellar masses. However, it should be emphasized that this can be done only
for a very limited number of systems, since the constraint $i \sim 90$\degr ~is quite strong.

In  
close binaries, due to the proximity of the companion the stars are no longer spherical.
This leads to the occurrence of finite quadrupolar, and higher,
momenta of the gravitational field (cf. \citealt{sterne}). 
These momenta force eccentric orbits to precess --i.e. modifies the
position of the periastron (apside or apse)--. 
Additionally, General Relativity predicts a secular
apsidal motion,
which is independent of the classical contributions, but 
it is related again with the components masses (cf. \citealt{levi}). 
It has been shown that the knowledge of the apsidal motion rate (AMR) allows
to estimate absolute masses in massive close binary systems with unknown orbital
inclination even in the case of non-eclipsing binaries \citep[][B02]{benvenuto}. 
From a theoretical point of view, the AMR is strongly dependent on the radii and internal structure 
of the stars of the pair. So, the masses determined by this method will be model dependent. 
In any case, accurate RV measurements and improved
stellar evolutionary models make 
currently feasible its use in reliable mass determinations.
In order to increase the number of O-type stars with determined absolute masses
we are conducting a systematic high-resolution spectroscopic monitoring of
binary systems with eccentric orbits with the aim of calculating their masses taking advantage of this method.

One of the objects that we are studying is  HD~165052 
(= CD~-24~13864; $\alpha_{2000}=18^{h}05^{m}10.6^{s}$, $\delta_{2000}=-24$\degr23'55'';
$V=6.87$)
an O-type double-lined spectroscopic binary star member of the open cluster NGC~6530 
in the \mbox{H\,{\sc ii}} region M8 (the Lagoon Nebula). 
It was first catalogued in the Cordoba Durchmusterung by \citet{thome}. 
\citet{plaskett1924} discovered variations in the RV of this star,
while double lines in its spectrum  were first noticed by \citet{conti74}.
\citet{morrison} presented the first orbital solution for the system (see Table \ref{tab:orb}).  
An improved circular orbital solution was calculated by \citet[][S97]{stickland} adding RV measurements from \textit{IUE}
spectra. A few years later, \citet[][A02]{arias} analysed high-resolution optical spectra, and 
determined a slightly eccentric ($e=0.09$) orbital solution. 
They fitted the previous RV data with their own solution and found evidence of 
apsidal motion.
\citet[][L07]{linder} determined a new orbital solution for the system which agrees with the A02 one.

In order to establish (or rule out) the AMR in HD~165052 
we have systematically observed this star between 2008 and 2010, and gathered a set of high-resolution spectra
(see Sec.~\ref{sec:obs}). 
Using the technique of disentangling
(cf. \citealt{gonzalez}) we have measured the RVs and simultaneously obtained separated spectra of the two components 
of the binary (see Sec.~\ref{sec:dis}). 
This procedure allowed us to re-classify the spectra of both components (Sec.~\ref{sec:spec}) 
and to fit a new orbital solution (Sec.~\ref{sec:orb}).
We have confirmed the existence of apsidal motion and measured the AMR for the first time (Sec.~\ref{sec:aps}). 
Then, applying the method mentioned above (B02) we have calculated the masses 
of the system components (Sec. \ref{sec:mas}).

\section{Observations}
\label{sec:obs}

We have acquired 37 high-resolution spectra of HD~165052 (see journal of observations on Table \ref{tab:journal}): 
29 at CASLEO, 
Argentina, with the 2.15-m Jorge Sahade telescope between 2008 August and 2010 August
using the REOSC SEL\footnote{Spectrograph Echelle Li\`ege
(jointly built by REOSC and Li\`ege Observatory, and on a long term loan from the latter.)} 
Cassegrain spectrograph in cross-dispersion mode;
5 at Las Campanas Observatory (LCO), Chile, with the Ir\'en\'ee du Pont 2.5-m telescope using the Echelle
Spectrograph in May 2010; 
and 3 additional spectra using FEROS 
at ESO La Silla, Chile, with the 2.2-m telescope in 2009 May.
The spectra taken at LCO and ESO La Silla
belong to the ``OWN Survey'' program \citep{barba}.
The mean signal to noise ratio (SNR) of all these spectra is $\sim 200$.
Technical details of the instrumental configurations can be found in Table \ref{tab:inst}.
All the spectra were reduced using the standard routines of IRAF\footnote{IRAF
 is distributed by the National Optical Astronomy Observatories,
    which are operated by the Association of Universities for Research
    in Astronomy, Inc., under cooperative agreement with the National
    Science Foundation.}.

\begin{table}
 \caption{Technical characteristics of the instruments in the configurations used for this work.}
\label{tab:inst}
 \begin{tabular}{@{}l ccc}
  \hline
Telescope -  instrument & reciprocal   & R        & spectral    \\
                        & dispersion   & @5000\AA & range  \\
                        & (\AA{}/pixel)&          & (\AA{})        \\
  \hline
CASLEO-J.S.-REOSC     & 0.19         &  13000 & 3600 - 6100    \\
ESO-2.2-FEROS         & 0.03         &  45000 & 3600 - 9200    \\
LCO-du Pont-Echelle   & 0.05         &  35000 & 3500 - 10100   \\
  \hline
 \end{tabular}

\end{table}

At CASLEO every night we have taken at least one spectrum of the star HD~137066, a RV standard established using 
CORAVEL photoelectric
scanner by \citet{andersen1985} with $v_r=-8.72 \pm 0.15$ km s$^{-1}$. 
We have also observed the B0V star $\tau$ Sco, 
considered as a rotational velocity standard
in the \citet{slettebak} system with $v \sin i < 10$ km s$^{-1}$. 
The latter spectra were taken with the same instrumental configuration as those of HD~165052
and averaged to achieve a SNR $\ga 600$ in the region of interest.

Spectra from CASLEO and LCO were calibrated using comparison spectra taken after every single exposition
from a Th-Ar lamp.

\section{Results and discussion}

\subsection{Disentangling and radial velocity measurements}
\label{sec:dis}

In order to measure the RVs, we applied the disentangling method
described by \citet{gonzalez} which allows to separate simultaneously the 
spectra of the binary components.
This iterative procedure consists in using alternatively the spectrum of one component to calculate 
the spectrum of the other one. In each step, the spectrum of one star is used to subtract its spectral 
features from the composite one. Then, the RVs are measured in the resulting single-lined spectra.
This spectra are shifted appropriately and combined to compute the new spectrum of each component. 

We have started the iterations using a template spectrum of the secondary star 
 taken from the grid of stellar atmosphere models 
in the \textsc{TLUSTY} database \citep{lanz} with solar metallicity\footnote{From here on subscripts 
1 and 2 will identify the stellar parameters of primary and secondary component of the system, respectively.}, 
$T_{\mathrm{eff},2} = 35000$K
and a $\log{g}_2=4.0$ that corresponds to main sequence star 
(cf. Sec. \ref{sec:spec}). The $T_{\mathrm{eff}}$ was chosen adopting 
the spectral type calibration of \citet{martins2005}.
The template was convolved to the projected rotational velocity
estimated by \citet{morrison}. As it has been shown by \cite{gonzalez}
these initial guesses do not affect the final results.   

The method also requires a RV initial value for each spectrum. A first raw measurement of the central 
wavelengths of the spectral lines \mbox{He\,{\sc i}} $\lambda\lambda$ 4471, 4921, 5876, 
\mbox{He\,{\sc ii}} $\lambda\lambda$ 4686 and 5411   
was made using \textsc{splot} task of IRAF. The RVs calculated from these lines were averaged to obtain
a first estimate of the RV of each component.  
Instead, during disentangling iterations the cross-correlation function was calculated over a disjoint sample region 
composed by 17 wavelengths intervals, each $\sim$10 \AA ~wide, taken around the following
spectral lines: \mbox{He\,{\sc i}} $\lambda\lambda$ 3819, 4026, 4387, 4471, 4713, 4921, 5015, 5875, 
\mbox{He\,{\sc ii}} $\lambda\lambda$ 4200, 4542, 4686, 5411,
\mbox{C\,{\sc iv}} $\lambda\lambda$ 5801, 5812,
\mbox{O\,{\sc iii}} $\lambda$ 5592,
\mbox{Mg\,{\sc ii}} $\lambda$ 4481 and
\mbox{Si\,{\sc iv}} $\lambda$ 4088.
The RVs thereby measured are listed in Table~\ref{tab:journal} and reproduced as \emph{mean} RVs in the 
Table~\ref{tab:vr}. \footnote{Table~\ref{tab:vr} is also available in electronic form at the Centre
de Données astronomiques de Strasbourg (CDS) 
http://webviz.u-strasbg.fr/viz-bin/VizieR.}

In order to provide detailed data for future works,  we have  subsequently 
defined  several sampling regions, each one around a single spectral line,  
and compute the cross-correlation separately for each region. So we have
measured the RVs of the individual lines listed in Table \ref{tab:vr}. 
The spectral lines selected were the same of A02 to allow for 
a comparison with that work. 
As it can be seen in Appendix \ref{lines}, the overall behavior of the RVs 
of the individual lines agrees with that of the \textit{mean} RV.

\begin{table}
\caption{Journal of the observations of HD~165052. Radial velocity measurements used to compute the orbital solution given in Table \ref{tab:orb}.}
\label{tab:journal}
 \begin{tabular}{@{}rcrrrrl}
  \hline
HJD             &phase  &$v_1$    &O-C    &$v_2$    &O-C& Obs.\\
$-2450000$      &$\phi$ &         &       &         &   & \\
\hline
4582.8675	&0.90	&97.1	&-2.6	&-108.8	&-2.4	&CAS\\
4693.6231	&0.38	&-81.6	&3.2	&91.7	&-3.9	&CAS\\
4695.6483	&0.06	&13.8	&0.9	&-8.8	&2.6	&CAS*\\
4696.5568	&0.37	&-85.4	&0.8	&95.1	&-1.9	&CAS\\
4696.6286	&0.39	&-80.2	&0.9	&94.3	&2.8	&CAS\\
4696.6605	&0.40	&-76.8	&1.6	&91.4	&2.8	&CAS\\
4697.6220	&0.73	&73.7	&2.9	&-76.1	&-1.4	&CAS\\
4955.8004	&0.10	&-7.6	&5.3	&5.1	&-11.7	&ESO*\\
4956.9100	&0.47	&-57.5	&-2.4	&61.0	&-2.0	&ESO\\
4964.8865	&0.17	&-57.6	&0.5	&66.7	&0.4	&CAS\\
4965.8323	&0.49	&-47.1	&0.0	&57.4	&3.2	&CAS\\
4966.8550	&0.84	&100.8	&-1.4	&-106.1	&2.9	&CAS\\
4967.8220	&0.16	&-56.1	&-1.3	&62.9	&0.2	&CAS\\
4968.8723	&0.52	&-38.8	&-4.6	&40.5	&0.5	&CAS\\
5046.6533	&0.84	&103.7	&1.3	&-109.9	&-0.5	&CAS\\
5047.7305	&0.21	&-73.8	&-0.7	&81.8	&-1.0	&CAS\\
5048.7185	&0.54	&-24.6	&-0.2	&29.3	&-0.2	&CAS*\\
5049.7159	&0.88	&106.4	&4.3	&-106.5	&2.5	&CAS\\
5052.7320	&0.90	&100.6	&1.5	&-104.9	&0.8	&CAS\\
5337.6475	&0.31	&-94.3	&-2.8	&101.3	&-1.6	&LCO\\
5339.7307	&0.02	&38.0	&-3.8	&-45.7	&-2.7	&LCO\\
5340.6327	&0.33	&-94.6	&-3.4	&100.7	&-1.8	&LCO\\
5341.5812	&0.65	&26.4	&-3.8	&-27.4	&2.9	&LCO*\\
5342.6292	&0.00	&54.0	&-0.5	&-60.9	&-4.0	&LCO\\
5376.5522	&0.48	&-52.2	&-0.3	&59.9	&0.4	&CAS\\
5378.7551	&0.23	&-79.7	&0.0	&88.5	&-1.4	&CAS\\
5380.5253	&0.82	&97.3	&-3.3	&-106.1	&1.2	&CAS\\
5381.7473	&0.24	&-84.7	&-1.7	&93.2	&-0.4	&CAS\\
5383.7515	&0.92	&91.5	&-3.2	&-98.5	&2.5	&CAS\\
5429.6974	&0.46	&-57.0	&1.1	&70.1	&3.8	&CAS\\
5430.6168	&0.78	&87.8	&-1.1	&-91.7	&2.9	&CAS\\
5431.6959	&0.14	&-42.6	&-1.1	&51.4	&3.2	&CAS\\
5432.6362	&0.46	&-58.1	&2.1	&69.5	&0.9	&CAS\\
5433.5059	&0.75	&81.0	&0.0	&-88.3	&-2.4	&CAS\\
5434.6834	&0.15	&-46.7	&1.2	&61.2	&6.1	&CAS\\
5435.5627	&0.45	&-65.0	&-1.2	&74.0	&1.4	&CAS\\
5698.8397	&0.54	&-6.0	&17.1	&41.2	&13.2	&ESO*\\
  \hline
\end{tabular}

\medskip Velocities and O-C in km s$^{-1}$. *: one tenth weight 
assigned in orbital solution fit because of strong lines blending.
Obs.: observatory where the spectrum was acquired.
CAS: CASLEO; ESO: ESO La Silla; LCO: Las Campanas.
\end{table}

\subsection{Spectral analysis}
\label{sec:spec}

We have visually compared our disentangled spectra
with those of the atlas of spectral standards published by \citet{sota} using the \textsc{MGB} code
\citep{maiz}.
We have observed in both components that the intensity of the absorption line  
\mbox{He\,{\sc ii}} $\lambda$ 4686 is greater than both \mbox{He\,{\sc i}} $\lambda$ 4471 and 
\mbox{He\,{\sc ii}} $\lambda$ 4542 (see Fig. \ref{fig:aspec}), a fact that 
is noted with a \textit{z} qualifier of the spectral type (cf. \citealt{walborn07}). 
Thus we classified the primary as O7V\textit{z} and the secondary as O7.5V\textit{z}.

In the past, other authors have classified these stars spectroscopically.
\citet{morrison} observed that
both stars are normal, with no conspicuous mass loss, and not far from the 
ZAMS. A02 showed that spectral types are O6.5V for the primary
 component and O7.5V for secondary. L07 concluded that the system
could be a O6.5V + O7V, or O6.5V + O7.5V, perhaps O((f)) because they detected 
\mbox{N\,{\sc iii}} $\lambda\lambda$ 4634, 40 and 41 in emission.
We see some traces of a very weak emission around 4640 in a couple of
FEROS spectra. 
In our disentangled primary-component spectrum, the absorption lines
\mbox{He\,{\sc i}} $\lambda$ 4471 and \mbox{He\,{\sc ii}} $\lambda$ 4542 seems to have
almost the same intensity. That is why we suspect it could be an O7V rather
than an O6.5V star. For the classification of the secondary, we agree with
the spectral type proposed by A02 and L07.

\subsection{Projected rotational velocity}
\label{sec:rotvel}

One of the stellar parameters required to mass determination is  
the projected rotational velocity $v \sin i$ of the binary components, since it is used
to compute the internal structure constants (see Sec.~\ref{sec:mas}).

To estimate the $v \sin i$, the spectrum of $\tau$~Sco has been convolved with 
 rotation line profiles calculated for different projected rotational velocities.
The \mbox{He\,{\sc i}} $\lambda\lambda$ 4713 and 5015 absorption lines were selected for these
measurements because they are isolated in the spectra and their Stark broadening can be considered as 
negligible (cf. \citealt{dimitrijevic}).
The full width at half maximum of intensity (\emph{FWHM}) of these lines in the convolved 
spectra of $\tau$ Sco were 
measured and a linear relation between \emph{FWHM} and the $v \sin i$ was fitted
supposing a rotational velocity very lower than the critical one (cf. \citealt{collins1974}).
Specifically it was found
 $FW\!H\!M_{4713}=0.0228 \times v\sin i + 0.219$ and 
 $FW\!H\!M_{5015}=0.0239 \times v\sin i + 0.277$ \AA, into the interval 
$30 \le v \sin i \le 100$ km s$^{-1}$. 
These empirical regressions were used to convert the
\emph{FWHM} of each line measured in the disentangled spectra in a $v \sin i$ value. 

Thereby we have computed $v_1\sin i = 71 \pm 5$ km s$^{-1}$ and $v_2\sin i = 66 \pm 5$ km s$^{-1}$.
Where the errors were estimated as half the uncertainty in the projected rotational velocity of $\tau$ Sco.

Other authors have estimated the $v \sin i$ of these stars, i.e. 
from \emph{IUE} spectra, S97 have found $v_1 \sin i$ = 85$\pm$8 and $v_2 \sin i=80 \pm 6$ km s$^{-1}$;
\citet{morrison} have obtained from the \emph{FWHM} intensity of the individual profiles of \mbox{He\,{\sc i}} $\lambda$
4471,  $v_1 \sin i = 65 \pm 9$ and $v_2 \sin i = 69 \pm 4$ km s$^{-1}$;
and L07 determined $v_1 \sin i = 73 \pm 7$ and $v_2 \sin i =80 \pm 7$ km s$^{-1}$ using the profiles of the
disentangled lines \mbox{He\,{\sc ii}} $\lambda$ 4471, \mbox{He\,{\sc ii}} $\lambda$ 4542 and H$\beta$.
From the four independent determinations, it is clear that the projected rotation velocity of both stars 
is very similar (within errors). Our results compare better with the later ones. 

\subsection{Orbital solution}
\label{sec:orb}

\begin{figure}
 \includegraphics[width=84mm]{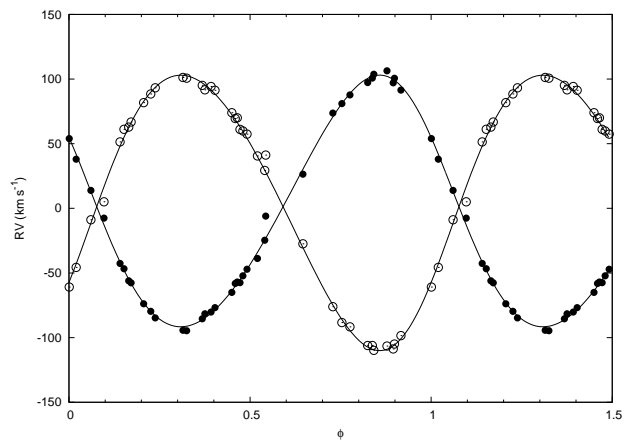}
\caption{Radial-velocity curve of HD~165052 calculated with the ephemerids
of the orbital solution found in this work (Table \ref{tab:orb}).
Measurements from Table \ref{tab:journal} are also represented 
(filled circles: primary radial velocities; open: secondary).
$\phi$ = 0 corresponds to the periastron passage.}
\label{fig:orb-sol}
\end{figure}

Our orbital solution was obtained using the RVs given in Table~\ref{tab:journal} as input for the \textsc{GBART}\footnote{Based on the 
algorithm of 
\citet{bertiau} and implemented by F. Bareilles
(available at http://www.iar.unlp.edu.ar/\textasciitilde fede/pub/gbart).} code.
The orbital parameters determined from the best-fitting are shown in the last column of 
Table~\ref{tab:orb} and depicted in Figure~\ref{fig:orb-sol}.
Our new orbital solution is close to the already published ones by S97, A02 and L07. 
The latter discussed about the
comparison among them. 
We also found a non-negligible eccentricity, thus confirming its reliability.
However, we point out the noticeable difference among the periastron longitudes ($\varpi$),
a clear sign of apsidal motion, which
will be analysed in the following.

\begin{table*}
 \begin{minipage}{150mm}
 \caption{Orbital solutions previously published for HD~165052  
to be compared with this work.}
\label{tab:orb}
 \begin{tabular}{@{}lccccc}
  \hline
Element             & M78                   & S97                  & A02                   & L07                   & this work        \\
  \hline
$P$ (days)          & $6.140 \pm 0.002$     & 2.955055             & $2.95510 \pm 0.00001$ & $2.95515 \pm 0.00004$ & $2.95506 \pm 0.00002$ \\ 
$e$                 & $0.064 \pm 0.041$     & 0.0 (assumed)        & $0.09 \pm 0.004$      & $0.081 \pm 0.015$     & $ 0.090 \pm 0.003$ \\
$\varpi$(\degr)     & $304 \pm 56$ (\dag)   & undefined            & $296.7 \pm 3.5$       & $298.0 \pm 10.2$      & $ 60 \pm 2 $\\
$T_0$               & $42939.5 \pm 0.6$     & $49819.075 \pm 0.008$& $49871.75 \pm 0.03$   & $51299.053 \pm 0.081$ & $55050.08 \pm 0.02$\\
(HJD-2400000)       &                       &                      &                       &                       & \\
$TV_{max}$          &                       &                      & $49872.19 \pm 0.03$   &                       & $55049.66 \pm 0.02$\\
(HJD-2400000)       &                       &                      &                       &                       & \\
$V_0$ (km s$^{-1}$) & $3.0 \pm 4.6$         & $-0.9 \pm 1.3$       & $1.05 \pm 0.31$       & $2.1 \pm 1.2$*        & $1.3 \pm 0.2$  \\
                    &                       &                      &                       &                       & \\
$K_1$ (km s$^{-1}$) & $91.0 \pm 2.8$        & $95.6 \pm 2.2$       & $94.8 \pm 0.5$        & $96.4 \pm 1.6$        & $97.4 \pm  0.4$ \\
$K_2$ (km s$^{-1}$) & $104.0 \pm 8.6$       & $109.6 \pm 2.2$      & $104.7 \pm 0.5$       & $113.5 \pm 1.9$       & $106.5 \pm 0.4$ \\
                    &                       &                      &                       &                       & \\
$a_1\sin i$ ($R_{\odot}$) & $11.1 \pm 0.3$  & $5.58 \pm 0.13$      & $5.51 \pm 0.03$       & $5.6 \pm 0.1$         & $5.66 \pm 0.03$\\
$a_2\sin i$ ($R_{\odot}$) & $12.7 \pm 0.7$  & $6.40 \pm 0.13$      & $6.09 \pm 0.03$       & $6.6 \pm 0.1$         & $6.20 \pm 0.03$ \\
                          &                 &                      &                       &                       & \\
$M_1 \sin^3 i$ ($M_{\odot}$) & $2.5 \pm 0.5$& $1.41 \pm 0.07$      & $1.26 \pm 0.03$       & $1.5 \pm 0.1$         & $1.34 \pm 0.03$ \\
$M_2 \sin^3 i$ ($M_{\odot}$) & $2.2 \pm 0.3$& $2.23 \pm 0.06$(\dag\dag)& $1.14 \pm 0.03$       & $1.3 \pm 0.1$         & $1.22 \pm 0.03$ \\
$q(M_2/M_1)$                 & $0.87 \pm 0.08$& $0.87 \pm 0.03$    & $0.90 \pm 0.01$       & $0.85 \pm 0.01$       & $0.91 \pm 0.01$ \\
                             &              &                      &                       &                       & \\
r.m.s. (km s$^{-1}$)         &              & 7.4                  & 2.21                  & 6.2                   & 1.7\\
  \hline
 \end{tabular}

 \medskip M78: \citet{morrison}; S97: \citet{stickland}; A02: \citet{arias}; 
L07: \citet{linder}. (\dag): measured from maximum positive radial velocity of star 1.
(\dag\dag): it seems that there was a typing mistake in this paper. The mass should probably be 1.23 $M_{\sun}$.
*: L07 fitted considering possible different systemic velocities for both components. 
Listed value corresponds to primary. For secondary they found $1.4 \pm 1.3$ km s$^{-1}$.

\end{minipage}
\end{table*}

\subsection{Apsidal motion}
\label{sec:aps}

\begin{figure}
 \includegraphics[width=84mm]{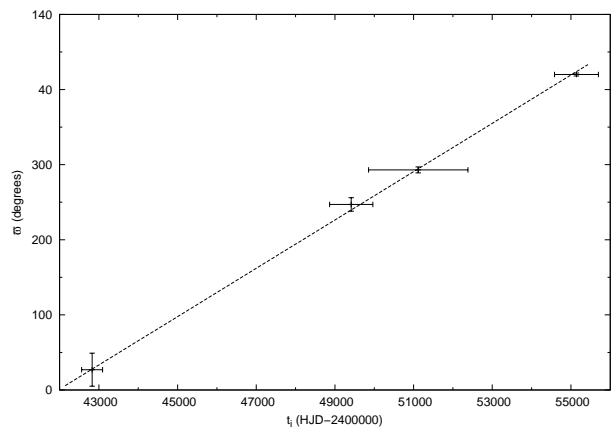}
\caption{Variation of the longitude of the periastron ($\varpi$) in time. Horizontal error bars
indicates the time span of each observational dataset. A linear fit with slope
$0.0322 \pm 0.0005$\degr day$^{-1}$ is shown  
(data from Table \ref{tab:omega}).}
\label{fig:omega}
\end{figure}

Our orbital solution (Table \ref{tab:journal}) gives $\varpi=60 \pm 2 \degr$, a very different value than 
those found by A02 and L07 --two solutions based on observations overlapped in time--. We
consider that this variation confirms the existence of apsidal motion in the system. To determine
its rate, we have re-computed  
orbital solutions, fixing $P=2.95506$~days and $e=0.090$  
(note that our values for these parameters do not differ significantly from those of S97, A02 and L07), to the previously 
published RVs. To do it we have grouped those data in four data sets according to the proximity of its 
observation date (see Table \ref{tab:omega}).
It means we have joined together the data from A02 and L07,  while we have 
not taken into account 4 data points from S97 between HJD 2444121.777 and 2445123.793.
It is worth note that the zero point in RV does not vary, within errors, between the different datasets.
We have considered each $\varpi_\mathrm{i}$ thus obtained as representative of the longitude of the periastron at an 
epoch $t_\mathrm{i}$ equal to the mid-point time between the first and the last date of the observations
in those particular dataset.  Since the variation of $\varpi$ in time 
seems to have a linear trend  (see Fig. \ref{fig:omega}) 
we have computed a linear regression, which slope was considered as a first
approximation to the AMR of the system:
\begin{equation}
 \dot{\varpi} = 0.0322 \pm 0.0005 \degr \mathrm{day}^{-1} .
\label{amrgbart} 
\end{equation}

 We have also used the \textsc{FOTEL} code developed by \citet{hadrava},
which allows to solve a RV curve taking into account the apsidal motion.
We have applied \textsc{FOTEL} to all the available RV (Table \ref{tab:rv-ant})
using as initial values our set of orbital parameters (Table \ref{tab:orb}, col. ``this work'')
and the $\dot{\varpi}$ value just found. 
Permitting the code to fit all the parameters simultaneously, the fitting process converged
to a set of orbital elements which agrees, within errors, with the values determined using the individual datasets.
The AMR thus obtained was
\begin{equation}
 \dot{\varpi} = 0.0340 \pm 0.0007 \degr \mathrm{day}^{-1} .
\label{amrfotel}
\end{equation}

\begin{table*}
\begin{minipage}{110mm}
 \caption{Longitude of the periastron $\varpi$ and others orbital elements at different epochs re-computed from 
previously published RVs measurements fixing $P=2.95506$ days and $e=0.090$.}
\label{tab:omega}
 \begin{tabular}{@{}lcccc}
  \hline
Dataset no.:        & 1                     & 2                     & 3                     & 4           \\
  \hline
initial date (HJD-2400000)    & 42560.940   & 48864.018             & 49854.6400            & 54582.8675  \\
final date (HJD-2400000)      & 43092.571   & 49965.110             & 52383.8577            & 55698.8397  \\
data from           & M78                   & S97                   & A02, L07              & this work        \\
                    &                       &                       &                       &             \\
$\varpi$(\degr)     & $27 \pm 22$           & $247 \pm 9$           & $293 \pm 4$           & $ 60 \pm 2 $\\
                    &                       &                       &                       &               \\                   
$V_0$ (km s$^{-1}$) & $0 \pm 2$             & $-1.7 \pm 0.9$        & $1.3 \pm 0.4$         & $1.3 \pm 0.2$  \\
                    &                       &                       &                       & \\
$K_1$ (km s$^{-1}$) & $94 \pm 4$            & $97 \pm 2$            & $95.4 \pm 0.8$        & $97.4 \pm 0.4$ \\
$K_2$ (km s$^{-1}$) & $103 \pm 4$           & $106 \pm 2$           & $107.8 \pm 0.8$       & $106.5 \pm 0.4$ \\
$q$                 & $0.91 \pm 0.07 $      & $0.91 \pm 0.03$       & $0.89 \pm 0.01$       & $0.91 \pm 0.01$\\
r.m.s. (km s$^{-1}$)& 8.4                   & 3.2                   & 4.1                   & 1.7 \\
  \hline
 \end{tabular}

 \medskip References as in Table \ref{tab:orb}. 
\end{minipage}
\end{table*}

In the following calculations we have used for AMR the value 
\begin{equation}
 \dot{\varpi} = 0.0331\pm 0.0009 \degr \mathrm{day}^{-1},
\end{equation}
since it is the mean between (\ref{amrgbart}) and (\ref{amrfotel}), while we have
adopted as its error the
semi-difference among them. (This AMR corresponds to $12.1 \pm 0.3 \degr$ yr$^{-1}$ 
or alternatively to an apsidal period $U = 29.8 \pm 0.8$ years).
It seems to be the highest AMR ever measured in an O+O system, 
with the possible exception of DH Cep (HD~215835) whose AMR has still to be
confirmed (cf. \citealt{petrova}, \citealt{bulut} and ref. therein).

\subsection{Calculation of the masses of the system employing the advance of the apside} \label{sec:teorico} 
\label{sec:mas}

As stated above (see Sec. \ref{intro}), the measurement of the secular advancement of the apside allows for 
the determination  of the masses of the components of the system even for the case of non-eclipsing binaries. The method to be employed below has been proposed and applied for the massive, non-eclipsing binary HD~93205 
by B02 (see also \citealt{jeffery}).

As it is well known, the gravitational potential of each component of the pair is affected by the 
presence of the other star and also by its own rotation. If we consider only the lowest (quadrupolar) 
correction to the gravitational potential 
of each object, the theoretical advancement of the apside is given by the Eq.~(14) of \citet{sterne}:

\begin{eqnarray}
 \frac{\dot{\varpi}}{\Omega_{\mathrm{orbit}}}= 
k_{2,1} \bigg(\frac{a_{1}}{A}\bigg)^5 
\bigg[ 15 \frac{M_{2}}{M_{1}} f_{2}(e) + \frac{\omega^2_1 A^3}{G M_{1}} g_{2}(e) \bigg] + \nonumber \\
k_{2,2} \bigg(\frac{a_{2}}{A}\bigg)^5 
\bigg[ 15 \frac{M_{1}}{M_{2}} f_{2}(e) + \frac{\omega^2_2 A^3}{G M_{2}} g_{2}(e) \bigg].
\label{eq:apsides} \end{eqnarray}

Here $\dot{\varpi}$ is the AMR; $\Omega_{\mathrm{orbit}}$ denotes the mean orbital angular velocity; 
$k_{\mathrm{i,j}}$ are the internal structure constants for the $i$-th multipolar term of the potential expansion 
(here $i=2$) and $j$ denotes the component of the pair (see below for details); $G$ is the gravitational constant; 
$A$ is the semi axis of the relative orbit; $M_{\mathrm{i}}$, $a_{\mathrm{i}}$, $\omega_{\mathrm{i}}$ are the mass, mean radius and 
angular rotation velocity of the $i$-th star respectively. $f_{2}(e)$ and $g_{2}(e)$ are functions of the 
eccentricity $e$ given by

\begin{equation}
f_2(e)= \frac{1 + \frac{3}{2} e^2 + \frac{3}{8} e^4}{(1-e^2)^5}
\end{equation} 

and

\begin{equation}
g_2(e)= \frac{1}{(1-e^2)^2}.
\end{equation}

The first and second terms of the r.h.s. of Eq.~(\ref{eq:apsides}) correspond to the contributions due to the primary and secondary stars, 
respectively. In each of these terms, the first term in the bracket is due to the tidal effect of one star on the other while 
the second one is due to stellar rotation.

\begin{equation}
\frac{\dot{\varpi}_{\mathrm{GR}}}{\Omega_{\mathrm{orbit}}}= 6.36 \times 10^{-6} \frac{M_{1} + M_{2}}{A(1-e^2)}, 
\end{equation} 

where masses and $A$ are in solar units. The internal structure constants $k_{\mathrm{i,j}}$ are dependent on the
structure of the stars. Most stellar evolution codes assume spherical symmetry, ignoring rotation. 
However, the departure 
from sphericity due to rotation modifies $k_{\mathrm{i,j}}$ appreciably. Fortunately, \citet{claret99} has shown 
that accounting for this correction is very simple, at least for $i=2$. If we define $[k_{2,\mathrm{j}}]_{\mathrm{sph}}$ as 
the internal structure constant corresponding to a spherical star, the corrected value of $k_\mathrm{{i,j}}$ is given by

\begin{equation}
\lg k_{2,\mathrm{j}}= \lg [k_{2,\mathrm{j}}]_{\mathrm{sph}} - 0.87 \frac{2 V^2_{\mathrm{j}}}{3 g_{\mathrm{j}} a_{\mathrm{j}}},
\end{equation} 

where $V_{\mathrm{j}}$ is the tangential velocity and $g_{\mathrm{j}}$ the surface gravitational acceleration. We shall consider 
 
\begin{equation}
\frac{\dot{\varpi}}{\Omega_{\mathrm{orbit}}} +
\frac{\dot{\varpi}_{\mathrm{GR}}}{\Omega_{\mathrm{orbit}}}=
\frac{\dot{\varpi}_{\mathrm{obs}}}{\Omega_{\mathrm{orbit}}} \label{eq:final} 
\end{equation} 

as an equation for $M_1$ as described in B02 where the reader will find further details on the method.
As discussed there, this method is model dependent, because $k_{\mathrm{i,j}}$ evolves due to changes in the 
density profile of the star and, more importantly, $a_{\mathrm{j}}$ also evolves. Notice the steep dependence of 
Eq.~(\ref{eq:apsides}) with the value of $a_{\mathrm{j}}$. Thus, as a matter of facts, this method is  
\textit{age dependent}. 

In order to apply the above described method, we consider that both stars have the same age and
are still burning 
hydrogen on their cores as it is indicated by  spectral classification.
We have computed solar composition stellar 
models with masses from 15~$M_{\odot}$ to 40~$M_{\odot}$ during core hydrogen burning in steps of 0.5~$M_{\odot}$. 
These models include mass loss as in \citet{jager} and overshooting as in \citet{demarque} the rest of the code 
corresponds to that described by \citet{benvenuto03} for binary evolution. 

\begin{figure}
\includegraphics[width=84mm]{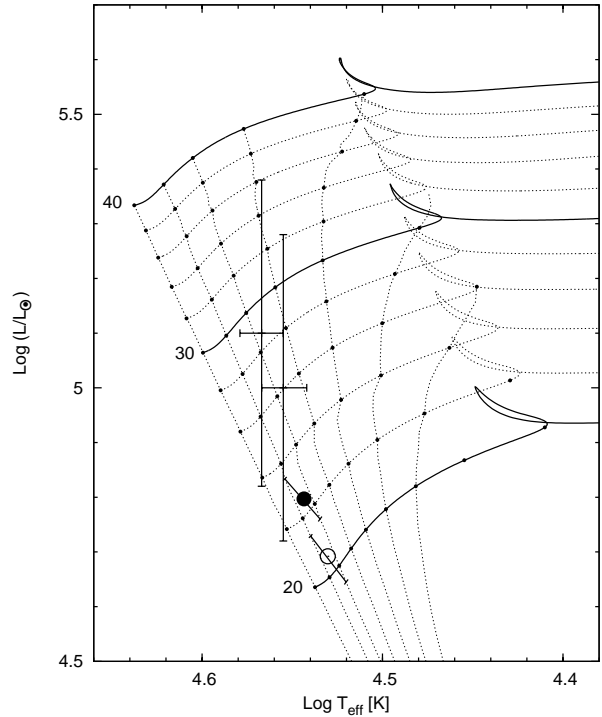}
\caption{\label{fig:hrd} Evolutionary tracks in the Hertzsprung - Russell diagram for stars of 
20, 30, and 40~$M_{\odot}$ (solid lines) 
and 22, 24, 26, 28, 32, 34, 36 and 38~$M_{\odot}$ (dotted lines) during core hydrogen burning. 
On each track, filled dots indicate time 
intervals of 1~Myr. Crosses: location of binary components according to literature. 
Horizontal error bars: $T_{\mathrm{eff}}$ dispersion in Martins et al. (2005, Table 4) observational
calibration. Vertical bars: lower limit: distance modulus from
\citet{mayne}. Upper limit: $M_V$ from \citet{buscombe} (see details in Sec. \ref{sec:age}).
 Circles: components derived from apsidal motion rate (filled primary);
error bars depict the age uncertainty propagated to masses.} 
\end{figure}

We show, in Fig.~\ref{fig:hrd} the Hertzsprung - Russell diagram for stellar masses in the range corresponding to the 
components of the pair with the evolutionary tracks calculated using these models. 
We set stellar ages to zero on the ZAMS.

\begin{figure}
\includegraphics[width=84mm]{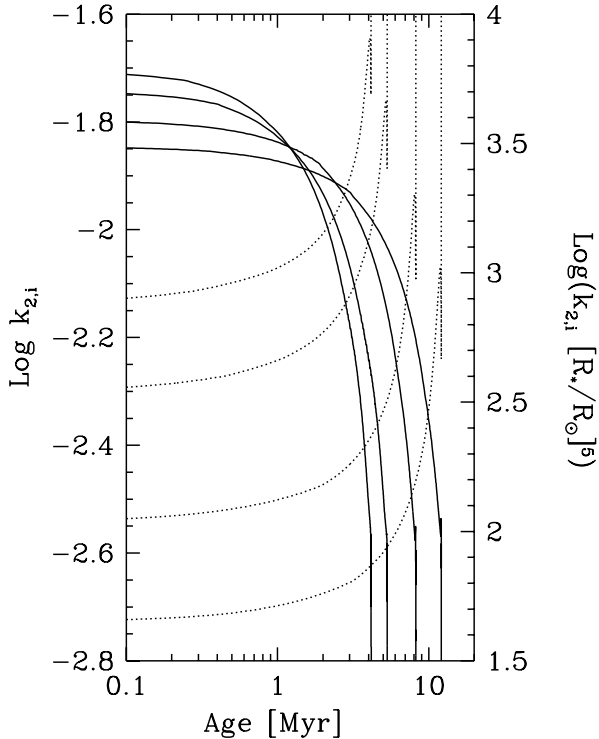}
\caption{Temporal evolution of the coefficients $k_{2,\mathrm{j}}$ (solid lines) and  $k_{2,\mathrm{j}}(R/R_{\odot})^5$ (dotted lines) 
for stars of 15, 20, 30, and 40~$M_{\odot}$. Curves end at larger ages the smaller is the mass. For further details 
see Sec. \ref{sec:mas}. \label{fig:k2} }
\end{figure}

\begin{figure}
\includegraphics[width=84mm]{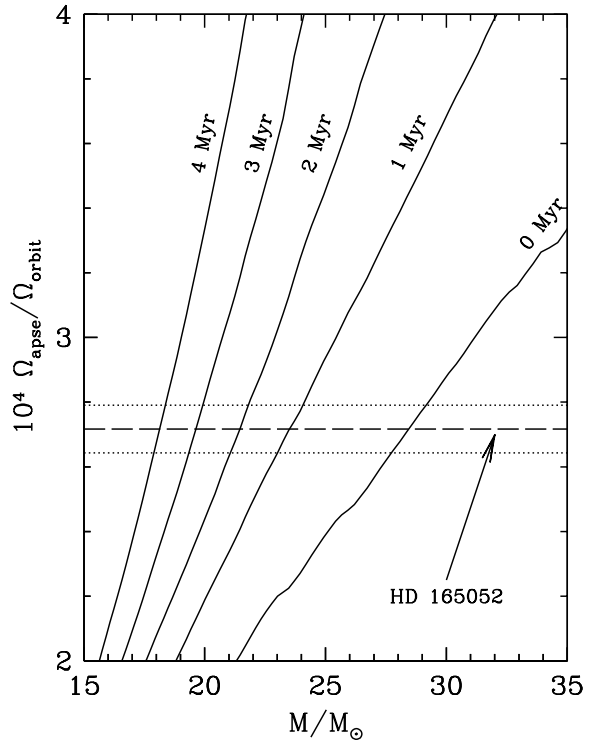}
\caption{The secular advance of the apse as a function of the mass of the primary component of HD~165052 
for different values of the age of the system, depicted in solid lines. The value corresponding to observations 
is denoted with an horizontal long dashed line. The value of the mass of the primary corresponds to the intersections: 
for ages of 0, 1, 2, 3, and 4~Myr the values for the mass of the primary are of 
$28.4 \pm 0.7$, $23.5 \pm 0.5$, 
$21.5 \pm 0.4$, $19.6 \pm 0.3$ and $18.1 \pm 0.3 M_{\sun}$ respectively. 
Evidently, the uncertainty in the age of the system has a direct impact on the determination of the masses 
of the components of the system. \label{fig:avanza} }
\end{figure}

Also, we have computed the coefficient $k_{2,\mathrm{j}}$ whose evolution, for different values of the initial mass is shown 
in Fig.~\ref{fig:k2} together with  $k_{2,\mathrm{j}}(R/R_{\odot})^5$, which appears in Eq.~(\ref{eq:apsides}). 
Now, we are in a position to employ the secular advancement of the apside ($\Omega_{\mathrm{apse}}\equiv \dot{\varpi}$) 
to determine the mass of the primary. 
A comparison of the theoretical results with observations is presented in Fig.~(\ref{fig:avanza}).  
 For a given age of the pair, the mass value of the primary corresponds to the intersection of the  
theoretical curve with the horizontal line at 
$\Omega_{\mathrm{apse}}/\Omega_{\mathrm{orbit}}= (2.72 \pm 0.08) \times 10^{-4}$.
The advancement of the apside is due 
to $\approx74$, $\approx23$ and $\approx 3$ per cent to tidal, rotational and relativistic contributions, respectively.

If we assume that the  
age of NGC~6530 is 1.5~Myr (see Sec. \ref{sec:age}) the most probable mass value for the primary 
of HD~165052 is $22.5 \pm 1.0~M_{\odot}$, 
where the error was
estimated considering an age uncertainty of 0.5~Myr. 
 Using the binary mass ratio $q$ from our orbital solution we found that the most probable
value for the mass of the secondary is $M_2= 20.5 \pm 0.9~M_{\sun}$ .

As HD~165052 is a much closer pair than HD~93205, it is worth to analyse the possibility of considering contributions to the apsidal motion due to terms beyond the quadrupolar. 
These contributions have been given by \citet{sterne}. It is found that the next term
in the expansion for the advancement of the apside is approximately
\begin{equation}\nonumber
2 \frac{k_{3,\mathrm{j}}}{k_{2,\mathrm{j}}} \bigg(\frac{a_{\mathrm{j}}}{A}\bigg)^2
\end{equation}
times the tidal term considered in Eq.~(\ref{eq:apsides}). Considering the solution given by Eq.~(\ref{eq:final}), 
we find that $a_{\mathrm{j}}/A\approx 1/3$ whereas the ratio of the internal structure constants can be estimated by employing 
the Table~15 of \citet{claret91} corresponding to a stellar model of 25~$M_{\odot}$. It is found that 
$k_{3,\mathrm{j}}/k_{2,\mathrm{j}}\approx 1/3$. Thus, the correction due to the first term beyond the quadrupolar one
is of the order of 4 per cent.
This is far smaller than the uncertainty in the theoretical AMR due to an error of 0.5~Myr in the age of 
the pair. This fact justifies the employment of Eq.~(\ref{eq:apsides}) up to the lowest order contribution.

The components masses that we have obtained are in agreement, within errors, with the spectroscopic masses derived 
by Martins et al. (2005) from a $T_{\mathrm{eff}}$ calibration of O stars. In fact, they obtained $25.3~M_{\sun}$ for an O7V star 
and $22.9~M_{\sun}$ for an O7.5V with an uncertainty as high as 35 to 50 per cent. 
On the other hand, the masses calculated are close to the range of masses most reliable determined via detached eclipsing binaries
for these spectral types. 
For example, the O7V stars V572~Car  (= Tr16~104), primary, and HD~165921 (V3903 Sgr), primary,  
have masses of $23.5 \pm 0.1~M_{\sun}$ and $27.27 \pm 0.55~M_{\sun}$ respectively (cf. \citealt{fernandezlajus} and \citealt{vaz}).
 
Nevertheless, the mass values we have derived from the apsidal motion rate 
are smaller than those estimated from photometric measurements and evolutionary tracks. 
We suspect that this difference should be mainly due 
to the large dispersion in the distance determinations of NGC~6530 found in literature, and probably also to the 
known mass discrepancy problem 
of O stars (cf. \citealt{massey}). Even so, when we plot in the H-R diagram (Fig. \ref{fig:hrd}) the points 
corresponding to our evolutionary models for
the age assumed and the masses calculated, we found that they are consistent with the luminosities calculated
from the most recent determinations of the cluster distance.

Additionally, the inclination of the orbit could be estimated from our mass determination. 
Taken the $M_1 \sin^3{i}$ value from our
orbital solution (Table \ref{tab:orb}) we derived $i \approx 23 \degr$, a result consistent with 
the fact that eclipses have never been reported.

The employed model is also capable of determining the radii of the components, giving $R_1\sim 7 R_{\sun}$ and $R_2\sim 6 R_{\sun}$.
These values are up to 30 per cent lower than those calibrated by Martins et al. (2005).
This fact has been already shown   
by \citet{fernandezlajus}.
He analysed a sample of detached eclipsing binaries whose components are O-type stars lying near the ZAMS, i.e.
V662~Car (= FO15 ), V572~Car, and V731~Car (= CPD-59~2635).
The same feature was found by \citet[][for HD~165921]{vaz} and by \citet[][for V573~Car = CPD-59~2628]{freyhammer}.

Furthermore, assuming this inclination and the radii of the components 
calculated with our models, 
from our orbital solution we obtain rotational periods of 2.0 and 1.8 days. It seems hence that
rotation is not synchronised with orbital motion ($P \approx 2.96$~d).
We plan to face this question in a further work, once that we have studied the apsidal
motion of our whole sample of systems.

Considering 
that the system is composed by two massive stars orbiting in a $\sim$3 day period, 
we explored the possibility of being in contact. Thus we calculated
$R_1/a_1=0.47$ and $R_2/a_2=0.40$ and compared them with the actual effective radii of their
Roche lobes $R_{L}$ \citep[using the formula given in][]{eggleton} which resulted 
$R_{L,1}/a_1=0.75$ and $R_{L,2}/a_2=0.67$. Therefore the stars are within their respective Roche lobes.

\subsection{On the age, distance and luminosity of HD~165052}
\label{sec:age}

As we have seen (Sec.~\ref{sec:mas}), the method described in B02 is particularly sensitive 
to the age of the binary system. In fact, the age of both stars is the parameter that introduces
the largest uncertainty in the calculation of the masses. Quite fortunately, HD~165052 
belongs to the open cluster NGC~6530. Thus, 
it is natural to consider that HD~165052 has the age of NGC~6530. In Table \ref{tab:age} we summarize
the age determinations found in the literature and the main related parameters. 

\citet{vanaltena} based on the theoretical gravitational contraction isochrones of \citet{iben} determined 
for the cluster an age of 2 Myr. \citet{kilambi} using $UBV$ photographic photometry estimated an age range of 
1-3 Myr for most of the stars 
in the gravitational contraction stage. \citet{bohm} studying $IUE$ spectra of the O-B stars and using theoretical evolutionary tracks inferred
an age of $\sim 5\pm 2$ Myr.
\citet{sung} obtained $UBVRI$ and $H_{\alpha}$ photometry of the cluster and comparing it with evolutionary models determined that the age of the most massive stars in the cluster is 1 - 2 Myr. \citet{damiani} 
identified X-ray sources in the cluster as 0.5-1.5 Myr age pre-main sequence stars with masses down to 
0.5-1.5 $M_{\sun}$ and found evidence of an age gradient from northwest to south. The median age of stars
in the central region of the cluster they have found was 0.8 Myr. \citet{prisinzano} presented $BVI$ photometry 
of the cluster and using evolutionary tracks found a median age of about 2.3 Myr. 
The most recent determination of the age of NGC~6530 is that of \citet{mayne} who gave a nominal value of 2~Myr fitting 
main-sequence models to data from literature sources.  
On the other hand, L07 calculated that the circularization time of this system should be less than 33500 years, 
but it is still eccentric; so, it should be very young (or its eccentricity is due to another physical process).

For our calculations we have assumed that the age of HD~165052 lies between 1 and 2 Myr. 
Thus, we have employed a 
value of 1.5 Myr with an error range of 0.5 Myr. This assumption is mainly supported by
the works of Sung et al. (2000) and \citet{mayne}.

It is worth mentioning that, apart from the works quoted above, other authors have determined the distance to NGC~6530 
(see \citealt[][Table 1]{prisinzano} and references therein) or the absolute magnitude of the system. 
Using these data, assuming the standard ratio of total to selective absorption
$R=3.1$ (whenever the authors did not estimated other value) 
and the bolometric corrections given by Martins et al. (2005, Table 4) 
for the binary components spectral types,
we have calculated the intrinsic luminosity of each component. To do it, we have assumed $V=6.87 \pm 0.01$, 
which is the mean value of the photometric measurements reported in SIMBAD database, and we have 
considered that the luminosity ratio $L_2/L_1$ could be taken from 
Martins et al. (2005, Table 4) observational calibration.

In this way, we have obtained $4.8 < \log (L_1/L_{\sun}) < 5.5$ 
The lower value corresponds to the distance modulus $m-M=10.34$ and color excess $E(B-V)=0.32$ determined by 
\citet{mayne}\footnote{Actually, \citet{mayne} reports in one of their fits (Table 8):
$10.15 < m-M < 10.44$ with a nominal value 10.34}. The upper value corresponds 
to $M_V=-4.8$ from \citet{buscombe}. These data are represented by the extremes of the 
vertical error bars in the Hertzsprung-Russell diagram in Fig. \ref{fig:hrd}. 
The horizontal error bars were traced at a luminosity level corresponding to the simple average of all the
published photometric data. For the secondary star we obtained $4.7 < \log (L_2/L_{\sun}) < 5.4$.
The large range in our luminosity estimations arises from the differences between the distance moduli
adopted by different authors.

In order to include the binary components in Fig. \ref{fig:hrd} we have assigned to each one the $T_{\mathrm{eff}}$
from the observational calibration of Martins et al. (2005, Table 4) for its spectral type. The horizontal bar
lengths indicates the dispersion calculated in the same work. 

\begin{table}
 \caption{Age determinations of NGC~6530.}
\label{tab:age}
 \begin{tabular}{@{}llclc}
  \hline
Ref. & $V_0 - M_v$     & $R$ & $\langle E(B-V)\rangle$ & Age  \\
     &                 &     &                         & (Myr) \\
\hline
VA72 &  11.25          & 3.0 & 0.35                    & 2 \\
K77  &  10.7           & 3.0 & 0.35 $\pm$ 0.01         & 1 - 3 \\
B84  & 11.5            & 3.2 &                         & $\sim 5 \pm 2$ \\
S00  & 11.25 $\pm$ 0.1 &     & 0.35                    & 1 - 2 \\
D04  &                 &     &                         & 0.8 \\
P05  & $\simeq$ 10.48  &     & 0.35                    & 2.3 \\
M08  & 10.15 - 10.44   &     & 0.32                    & 2 \\
\hline
\end{tabular}

\medskip $R$: ratio of total to selective absorption. References: VA72: \citet{vanaltena};
K77: \citet{kilambi}; B84: Boehm-Vitense et al (1984); S00: Sung et al. (2000); D04: \citet{damiani}; P05: \citet{prisinzano};
M08: \citet{mayne}.
\end{table}

\section{Summary}

We have observed the spectroscopic binary HD~165052 gathering a set of high-resolution and high-S/N spectra,
from which we have measured the radial velocity of its components over its orbital movement.
With these data we have determined the parameters of its current orbit confirming that it is eccentric,
as has been realized by \citet{arias}. Re-analysing together all the previously published radial velocity measurements,
we have
demonstrated the precession of the orbit, a fact suggested by A02. 
We have also determined by the first
time de apsidal motion rate of the system $\dot{\varpi} = 0.0331 \pm 0.0009\degr$day$^{-1}$
which seems to be the highest value ever measured in an O+O binary system.

We have disentangled the components spectra and re-classified it, founding that both present the
\textit{z} spectral feature.
Then, we classify the primary as 
O7V\textit{z} and the secondary as O7.5V\textit{z}.

Using the apsidal motion rate, with the method described in \citet{benvenuto}, we have calculated
the absolute masses of the binary components ($M_1=22.5 \pm 1.0~M_{\odot}$, 
$M_2=20.5 \pm 0.9~M_{\sun}$). 
These masses lies close to those determined for eclipsing binaries of the same spectral types and  
compare well with those theoretically estimated in the Martins et al. (2005) calibration.

We have estimated the luminosities of the binary components, but unfortunately there is a large uncertainty
in these calculations because of the large differences in the distances previously determined to the
cluster NGC~6530. This is why the masses estimated from photometry and evolutionary models are very
uncertain. Nonetheless, the masses that we have obtained from apsidal
motion rate suggested that the distance to the cluster could be around the smallest determined until now. 
To solve this apparent inconsistency should be necessary to have new independent determinations of the
distance to the cluster.

\section*{Acknowledgments}

We acknowledge our referee, Ian Howarth, for his very useful comments which improved substantially this paper.

We are very grateful to Nidia Morrell and Rodolfo Barb\'a because of their kind collaboration in the 
acquisition of the spectra at Las Campanas and ESO-La Silla observatories. We also acknowledge Petr Hadrava
for kindly allowing us to use the \textsc{FOTEL} code.

We thank the directors and staffs of CASLEO, LCO and ESO, La Silla, for the use of their facilities
and their gentle hospitality during the observing runs.

This research has made use of the NASA's Astrophysics Data System and the SIMBAD database,
operated at CDS, Strasbourg, France.

\clearpage

\appendix
\section{Disentangled spectra}
\label{spectra}

\begin{figure*}
\begin{minipage}{170mm}
 \includegraphics[width=\linewidth]{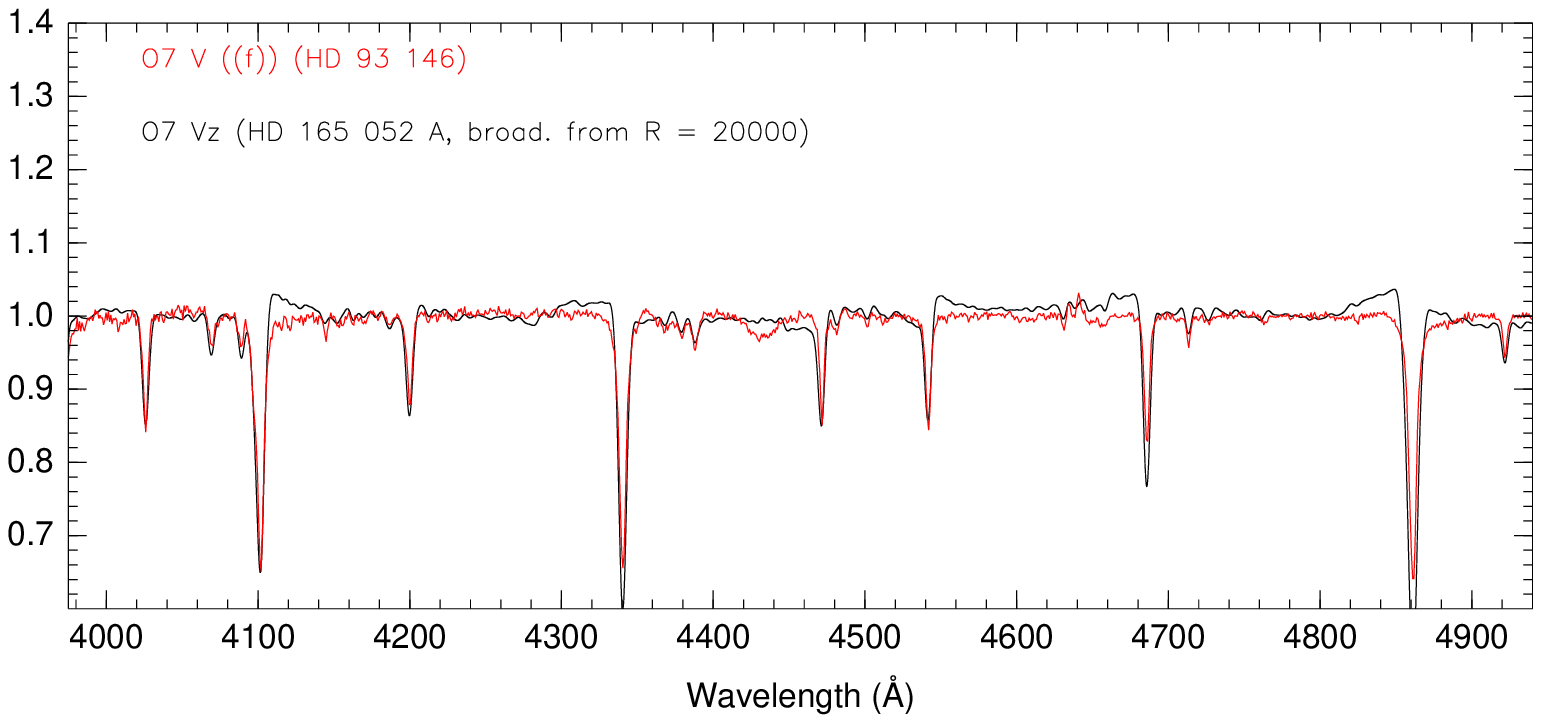}
 \includegraphics[width=\linewidth]{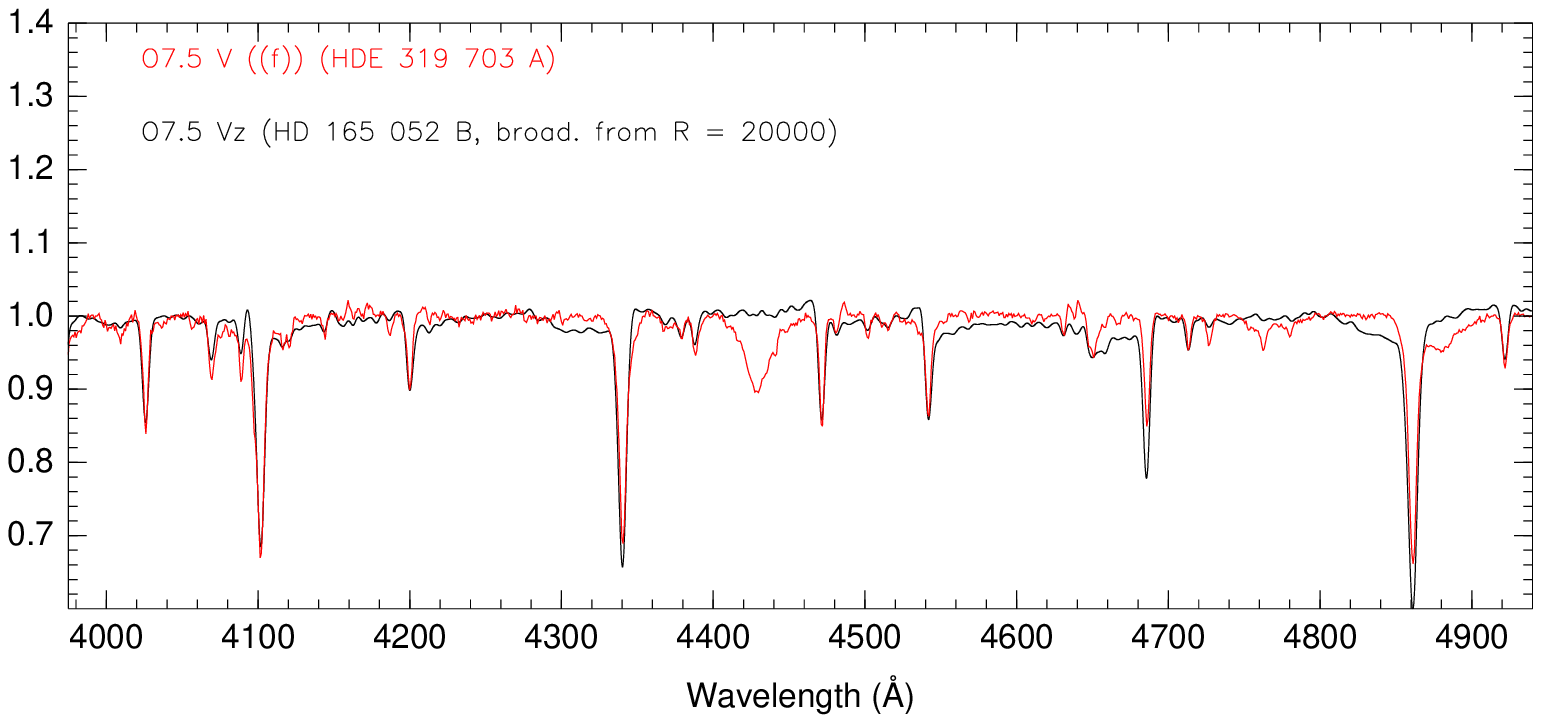}
\caption{HD~165052 primary (A) and secondary (B) star spectra resample to $R \sim 2500$ (black) compared with that of the
spectral standards HD~93146 and HDE~319703~A (red). 
It could be note the greater intensity of \mbox{He\,{\sc ii}} $\lambda$ 4686 
absorption line (the \textit{z} feature).
The difference between the continuum level at bluer and redder wing of some lines is due to an artefact of the
disentangling algorithm.
 Figures generated with \textsc{MGB} code \citep{sota}.}
\label{fig:aspec}
\end{minipage}
\end{figure*}

\section{Tables of radial velocities}

\subsection{Radial velocities from literature}

%\begin{longtable}{@{}lcccl}
%\caption{Radial velocities of HD~165052 previously reported.}
%\label{tab:rv-ant}
% \hline
%  HJD-2400000 & $v_{1}$ & $v_{s}$ & $v_{2}$ & Ref.        \\
%  \hline
%\endfirsthead
%\caption{Radial velocities of HD~165052 previously reported. (Continued)}
% \hline
%  HJD-2400000 & $v_{1}$ & $v_{s}$ & $v_{2}$ & Ref.        \\
%  \hline
%\endhead
%  \hline
%\endfoot
%  \hline
% \medskip
%$v_s$: systemic velocity.
%\textbf{Ref.}: H32: \citet{hayford};
%C77: \citet{conti77};
%L07: \citet{linder};
%M78: \citet{morrison};
%P24: \citet{plaskett1924}; 
%S49: \citet{sanford};
%S97: \citet{stickland};
%W53: \citet{wilson53}.
%\endlasfoot

\begin{table}
 \caption{Radial velocities of HD~165052 previously reported.}
\label{tab:rv-ant}
 \begin{tabular}{@{}lcccl}
  \hline
  HJD-2400000 & $v_{1}$ & $v_{s}$ & $v_{2}$ & Ref.        \\
  \hline

$\sim 24000$&    & $3.0$          &           & P24 \\
$25371.99$  &    & $6.0$          &           & H32 \\
$\sim 33000$&    & $-1.0 \pm 8.0$ &           & S49 \\      
$\sim 35000$&    & $3.0 \pm 2.5  $&           & W53 \\
40075.70  &      & $-6.6 \pm 3.2 $&           & C77 \\
40047.80  &      & $-3.1 \pm 4.5 $&           & C77 \\
          &      &                &           &     \\
42560.940 &      & -9 &      & M78 \\
42561.858 &      &  1 &      & M78 \\
42562.921 &  -73 &    &   93 & M78 \\
42565.822 & -106 &    &   95 & M78 \\
42566.850 &      &  1 &      & M78 \\
          &      &    &      &     \\
42687.587 &      & 12 &      & M78 \\
42688.588 &  103 &    & -121 & M78 \\
42689.591 &  -91 &    &   62 & M78 \\
42691.584 &  100 &    &  -94 & M78 \\
42937.771 &  -63 &    &   64 & M78 \\
          &      &    &      &     \\
42938.861 &      & -1 &      & M78 \\
42939.800 &  101 &    & -111 & M78 \\
42940.918 &  -81 &    &   88 & M78 \\
42941.851 &      &  5 &      & M78 \\
42942.902 &   74 &    &  -96 & M78 \\
          &      &    &      &     \\
42943.863 &  -61 &    &   91 & M78 \\
43089.575 &      & 11 &      & M78 \\
43090.573 &  110 &    & -105 & M78 \\
43091.579 &  -66 &    &  101 & M78 \\
43092.571 &      & 10 &      & M78 \\
          &      &    &      &     \\
44121.777 &   99.2 &  & -109.8 & S97 \\
44459.983 &  -83.4 &  &   98.4 & S97 \\
44899.113 &   90.9 &  & -105.5 & S97 \\
45123.793 &   89.6 &  &  -86.7 & S97 \\
48864.018 &   39.0 &  &  -42.9 & S97 \\
          &      &    &      &     \\
48864.053 &   47.0 &  &  -48.4 & S97 \\
48867.621 &  100.0 &  &  -91.8 & S97 \\
49101.414 &   54.7 &  &  -67.9 & S97 \\
49399.573 &   86.7 &  &  -99.5 & S97 \\
49817.685 & -100.1 &  &  106.9 & S97 \\
          &      &    &      &     \\
49818.419 &   13.5 &  &  -25.3 & S97 \\
49818.842 &   79.4 &  & -100.0 & S97 \\
49819.417 &   61.0 &  &  -79.5 & S97 \\
49821.120 &  -50.2 &  &   38.6 & S97 \\
49965.110 &  -72.5 &  &   87.8 & S97 \\
          &      &    &      &     \\
49854.640 &   88.2 &  &  -98.5 & A02 \\
49931.634 &   69.2 &  &  -91.4 & A02 \\
49932.663 &  -76.5 &  &   95.4 & A02 \\
49934.577 &   75.2 &  &  -82.8 & A02 \\
49935.605 &  -77.6 &  &   88.5 & A02 \\
          &      &    &      &     \\
49937.587 &   71.9 &  &  -73.2 & A02 \\
50239.906 &  -74.0 &  &   76.8 & A02 \\
50241.823 &   82.5 &  &  -90.9 & A02 \\
50244.902 &   72.6 &  &  -76.5 & A02 \\
50245.857 &  -74.7 &  &   82.5 & A02 \\
          &      &    &      &     \\
50293.763 &  -74.5 &  &   82.6 & A02 \\
50296.787 &  -63.2 &  &   64.3 & A02 \\
  \hline
 \end{tabular}
\end{table}

\newpage
\setcounter{table}{0}
\begin{table}
 \caption{Continued.}
\label{tab:rv-ant2}
 \begin{tabular}{@{}lcccl}
  \hline
  HJD-2400000 & $v_{1}$ & $v_{s}$ & $v_{2}$ & Ref.        \\
  \hline
50671.471 &  -78.1 &  &   94.0 & A02 \\
51299.7250 &  88.3 &  & -100.8  & L07 \\
51300.7319 & -56.5 &  &   76.9  & L07 \\
51300.9264 & -86.7 &  &   96.6  & L07 \\
51301.9281 &  31.2 &  &  -24.9  & L07 \\
           &       &  &         &     \\
51304.7434 & 7.4  &  &   -1.7  & L07 \\
51304.7507 &  15.0 &  &  -12.5  & L07 \\
51304.9309 & 40.1 &  &  -39.5  & L07 \\
51323.8361 & 17.9 &  &  -25.8  & L07 \\
51327.6014 & -87.7 &  &  106.8    & L07 \\
           &       &  &         &     \\
51327.9127 &  -93.7 &  &  103.3  & L07 \\
51670.7601 &  -80.1 &  &  103.7  & L07 \\
51671.7225 &  96.3 &  & -111.3  & L07 \\
51672.7016 & -15.9 &  &   28.7  & L07 \\
51714.835 &  -87.2 &  &  102.9 & A02 \\
           &       &  &         &     \\
51714.862 &  -93.5 &  &   99.2 & A02 \\
51715.796 &   61.2 &  &  -62.9 & A02 \\
51716.570 &   71.6 &  &  -68.5 & A02 \\
51716.656 &   34.0 &  &  -41.9 & A02 \\
51717.609 &  -80.7 &  &   95.6 & A02 \\
           &       &  &         &     \\
51717.727 &  -90.5 &  &   95.3 & A02 \\
51717.829 &  -93.0 &  &   97.7 & A02 \\
52066.791 &  -70.1 &  &   90.6 & A02 \\
52067.768 &   95.6 &  & -110.5 & A02 \\
52069.730 &  -75.4 &  &   86.6 & A02 \\
           &       &  &         &     \\
52069.794 &  -67.3 &  &   76.2 & A02 \\
52070.621 &   93.3 &  &  -97.1 & A02 \\
52070.731 &  103.5 &  & -104.9 & A02 \\
52070.792 &  102.8 &  & -107.4 & A02 \\
52072.663 &  -75.3 &  &   90.1 & A02 \\
           &       &  &         &     \\
52072.717 &  -69.4 &  &   78.9 & A02 \\
52335.8879 & -41.0 &  &   56.9  & L07 \\
52336.8791 &  97.5 &  & -118.5  & L07 \\
52337.8880 & -54.9 &  &   70.1  & L07 \\
52338.8808 & -46.1 &  &   47.8  & L07 \\
           &       &  &         &     \\
52339.8848 &  96.9 &  & -111.2  & L07 \\
52381.8324 &   6.6 &  &   -5.7  & L07 \\
52382.8569 & -79.7 &  &  103.3   & L07 \\
52383.8577 &  95.4 &  & -103.1  & L07 \\
  \hline
 \end{tabular}

 \medskip
$v_s$: systemic velocity.
\textbf{Ref.}: H32: \citet{hayford};
C77: \citet{conti77};
L07: \citet{linder};
M78: \citet{morrison};
P24: \citet{plaskett1924}; 
S49: \citet{sanford};
S97: \citet{stickland};
W53: \citet{wilson53}.
\end{table}
%
%\end{longtable}

\subsection{Individual lines radial-velocity measurements}
\label{lines}

\setcounter{table}{1}
\begin{table*}
\vbox to220mm{\vfil Landscape table with radial velocities measurements from individual lines of HD~165052 to go here. 
\caption{}
\vfil}
\label{tab:vr}
\end{table*}

\bsp

\label{lastpage}


\begin{thebibliography}{}

\bibitem[\protect\citeauthoryear{Andersen et 
al.}{1985}]{andersen1985} Andersen J., et al., 1985, A\&AS, 59, 15 

\bibitem[\protect\citeauthoryear{Arias et al.}{2002}]{arias} 
Arias J.~I., Morrell N.~I., Barb{\'a} R.~H., Bosch G.~L., Grosso M., 
Corcoran M., 2002, MNRAS, 333, 202 (A02)

\bibitem[\protect\citeauthoryear{Barb{\'a} et 
al.}{2010}]{barba} Barb{\'a} R.~H., Gamen R., Arias J.~I., 
Morrell N., Ma{\'{\i}}z Apell{\'a}niz J., Alfaro E., Walborn N., Sota A., 
2010, RMxAC, 38, 30 

\bibitem[\protect\citeauthoryear{Benvenuto 
\& De Vito}{2003}]{benvenuto03} Benvenuto O.~G., De Vito M.~A., 2003, MNRAS, 342, 50 

\bibitem[\protect\citeauthoryear{Benvenuto et 
al.}{2002}]{benvenuto} Benvenuto O.~G., Serenelli A.~M., Althaus 
L.~G., Barb{\'a} R.~H., Morrell N.~I., 2002, MNRAS, 330, 435 (B02) 

\bibitem[\protect\citeauthoryear{Bertiau 
\& Grobben}{1969}]{bertiau} Bertiau F.~C., Grobben J., 1969, Ric. Astron. Sp. Vaticana, 8, 1 

\bibitem[\protect\citeauthoryear{Boehm-Vitense, Hodge, 
\& Boggs}{1984}]{bohm} Boehm-Vitense E., Hodge P., Boggs D., 1984, ApJ, 287, 825 

\bibitem[\protect\citeauthoryear{Bulut 
\& Demircan}{2007}]{bulut} Bulut I., Demircan O., 2007, MNRAS, 378, 179 

\bibitem[\protect\citeauthoryear{Buscombe}{1969}]{buscombe} 
Buscombe W., 1969, MNRAS, 144, 31 

\bibitem[\protect\citeauthoryear{Claret}{1999}]{claret99} Claret A., 1999, A\&A, 350, 56 

\bibitem[\protect\citeauthoryear{Claret 
\& Gimenez}{1991}]{claret91} Claret A., Gimenez A., 1991, A\&AS, 87, 507 

\bibitem[\protect\citeauthoryear{Collins}{1974}]{collins1974} 
Collins G.~W., II, 1974, ApJ, 191, 157 

\bibitem[\protect\citeauthoryear{Conti}{1974}]{conti74} Conti 
P.~S., 1974, ApJ, 187, 539 

\bibitem[\protect\citeauthoryear{Conti, Leep, 
\& Lorre}{1977}]{conti77} Conti P.~S., Leep E.~M., Lorre J.~J., 1977, ApJ, 214, 759 

\bibitem[\protect\citeauthoryear{Damiani et 
al.}{2004}]{damiani} Damiani F., Flaccomio E., Micela G., 
Sciortino S., Harnden F.~R., Jr., Murray S.~S., 2004, ApJ, 608, 781 

\bibitem[\protect\citeauthoryear{de Jager, Nieuwenhuijzen, 
\& van der Hucht}{1988}]{jager} de Jager C., Nieuwenhuijzen H., van der Hucht K.~A., 1988, A\&AS, 72, 259 

\bibitem[\protect\citeauthoryear{Demarque et 
al.}{2004}]{demarque} Demarque P., Woo J.-H., Kim Y.-C., Yi 
S.~K., 2004, ApJS, 155, 667 

\bibitem[\protect\citeauthoryear{Dimitrijevic 
\& Sahal-Brechot}{1990}]{dimitrijevic} Dimitrijevic M.~S., Sahal-Brechot S., 1990, A\&AS, 82, 519 

\bibitem[\protect\citeauthoryear{Eggleton}{1983}]{eggleton} 
Eggleton P.~P., 1983, ApJ, 268, 368 

\bibitem[\protect\citeauthoryear{Fern\'andez Laj\'us}{2006}]{fernandezlajus} Fern\'andez Laj\'us E., 2006, PhD thesis, Univ.
Nac. de La Plata, Argentina

\bibitem[\protect\citeauthoryear{Freyhammer et 
al.}{2001}]{freyhammer} Freyhammer L.~M., Clausen J.~V., Arentoft T., Sterken C., 2001, A\&A, 369, 561 

\bibitem[\protect\citeauthoryear{Gonz{\'a}lez 
\& Levato}{2006}]{gonzalez} Gonz{\'a}lez J.~F., Levato H., 2006, A\&A, 448, 283 

\bibitem[\protect\citeauthoryear{Hadrava}{2004}]{hadrava} 
Hadrava P., 2004, Publ. Astron. Inst. Acad. Sci. Czech Rep., 92, 1 

\bibitem[\protect\citeauthoryear{Hayford}{1932}]{hayford} 
Hayford P., 1932, Lick Obs. Bull., 16, 53 

\bibitem[\protect\citeauthoryear{Iben}{1965}]{iben} Iben I., 
Jr., 1965, ApJ, 141, 993 

\bibitem[\protect\citeauthoryear{Jeffery}{1984}]{jeffery} 
Jeffery C.~S., 1984, MNRAS, 207, 323 

\bibitem[\protect\citeauthoryear{Kilambi}{1977}]{kilambi} 
Kilambi G.~C., 1977, MNRAS, 178, 423 

\bibitem[\protect\citeauthoryear{Lanz 
\& Hubeny}{2003}]{lanz} Lanz T., Hubeny I., 2003, ApJS, 146, 417 

\bibitem[\protect\citeauthoryear{Levi-Civita}{1937}]{levi} Levi-Civita, T., 1937, Am. J. Math., 59, 225

\bibitem[\protect\citeauthoryear{Linder et 
al.}{2007}]{linder} Linder N., Rauw G., Sana H., De Becker M., Gosset E., 2007, A\&A, 474, 193 (L07)

\bibitem[\protect\citeauthoryear{Ma{\'{\i}}z Apell{\'a}niz et 
al.}{2011}]{maiz} Ma{\'{\i}}z Apell{\'a}niz J., Sota A., 
Walborn N.~R., Alfaro E.~J., Barb{\'a} R.~H., Morrell N.~I., Gamen R.~C., 
Arias J.~I., 2011, Highlights of Spanish Astrophysics VI, 467 

\bibitem[\protect\citeauthoryear{Martins, Schaerer, 
\& Hillier}{2005}]{martins2005} Martins F., Schaerer D., Hillier D.~J., 2005, A\&A, 436, 1049 

\bibitem[\protect\citeauthoryear{Massey et al.}{2012}]{massey} 
Massey P., Morrell N.~I., Neugent K.~F., Penny L.~R., DeGioia-Eastwood K., 
Gies D.~R., 2012, ApJ, 748, 96 

\bibitem[\protect\citeauthoryear{Mayne 
\& Naylor}{2008}]{mayne} Mayne N.~J., Naylor T., 2008, MNRAS, 386, 261 

\bibitem[\protect\citeauthoryear{Morrison 
\& Conti}{1978}]{morrison} Morrison N.~D., Conti P.~S., 1978, ApJ, 224, 558 (M78)

\bibitem[\protect\citeauthoryear{Petrova 
\& Orlov}{1999}]{petrova} Petrova A.~V., Orlov V.~V., 1999, AJ, 117, 587 

\bibitem[\protect\citeauthoryear{Plaskett}{1924}]{plaskett1924} 
Plaskett J., 1924, Pub. DAO, 2, 286 

\bibitem[\protect\citeauthoryear{Prisinzano et 
al.}{2005}]{prisinzano} Prisinzano L., Damiani F., Micela G., Sciortino S., 2005, A\&A, 430, 941 

\bibitem[\protect\citeauthoryear{Sanford}{1949}]{sanford} 
Sanford R.~F., 1949, ApJ, 110, 117 

\bibitem[\protect\citeauthoryear{Slettebak et 
al.}{1975}]{slettebak} Slettebak A., Collins G.~W., II, Parkinson 
T.~D., Boyce P.~B., White N.~M., 1975, ApJS, 29, 137 

\bibitem[\protect\citeauthoryear{Sota et al.}{2011}]{sota} 
Sota A., Ma{\'{\i}}z Apell{\'a}niz J., Walborn N.~R., Alfaro E.~J., 
Barb{\'a} R.~H., Morrell N.~I., Gamen R.~C., Arias J.~I., 2011, ApJS, 193, 
24 

\bibitem[\protect\citeauthoryear{Sterne}{1939}]{sterne} Sterne 
T.~E., 1939, MNRAS, 99, 451 

\bibitem[\protect\citeauthoryear{Stickland, Lloyd, 
\& Koch}{1997}]{stickland} Stickland D.~J., Lloyd C., Koch R.~H., 1997, The Observatory, 117, 295 (S97)

\bibitem[\protect\citeauthoryear{Sung, Chun, 
\& Bessell}{2000}]{sung} Sung H., Chun M., Bessell M., 2000, AJ, 120, 333 

\bibitem[\protect\citeauthoryear{Thome}{1892}]{thome} Thome 
J.~M., 1892, Res. National Argentine Obs., 16, 1 

\bibitem[\protect\citeauthoryear{van Altena 
\& Jones}{1972}]{vanaltena} van Altena W.~F., Jones B.~F., 1972, A\&A, 20, 425 

\bibitem[\protect\citeauthoryear{Vaz et 
al.}{1997}]{vaz} Vaz L.~P.~R., Cunha N.~C.~S., Vieira E.~F., Myrrha M.~L.~M., 1997, A\&A, 327, 1094 

\bibitem[\protect\citeauthoryear{Walborn}{2009}]{walborn07} 
Walborn N.~R., 2009, STScI Symposium Series No. 20, 167 

\bibitem[\protect\citeauthoryear{Wilson}{1953}]{wilson53} Wilson 
R.~E., 1953, General Catalogue of Stellar Radial Velocities, Carnegie Inst. Washington D.C. Publ.

\end{thebibliography}
\end{document}